\documentclass[preprint,aps,superscriptaddress,nofootinbib,tightenlines]{revtex4}
\usepackage{epsfig}
\usepackage{bm}

\def\OMIT#1{{}}
\def\Dslash{D\hskip-0.65em /}
\newcommand{\gsim}{\raisebox{-0.7ex}{$\stackrel{\textstyle >}{\sim}$ }}
\newcommand{\lsim}{\raisebox{-0.7ex}{$\stackrel{\textstyle <}{\sim}$ }}

\begin{document}

\begin{figure}[!t]
\vskip -1.4cm
\leftline{
{\epsfxsize=1.5in \epsfbox{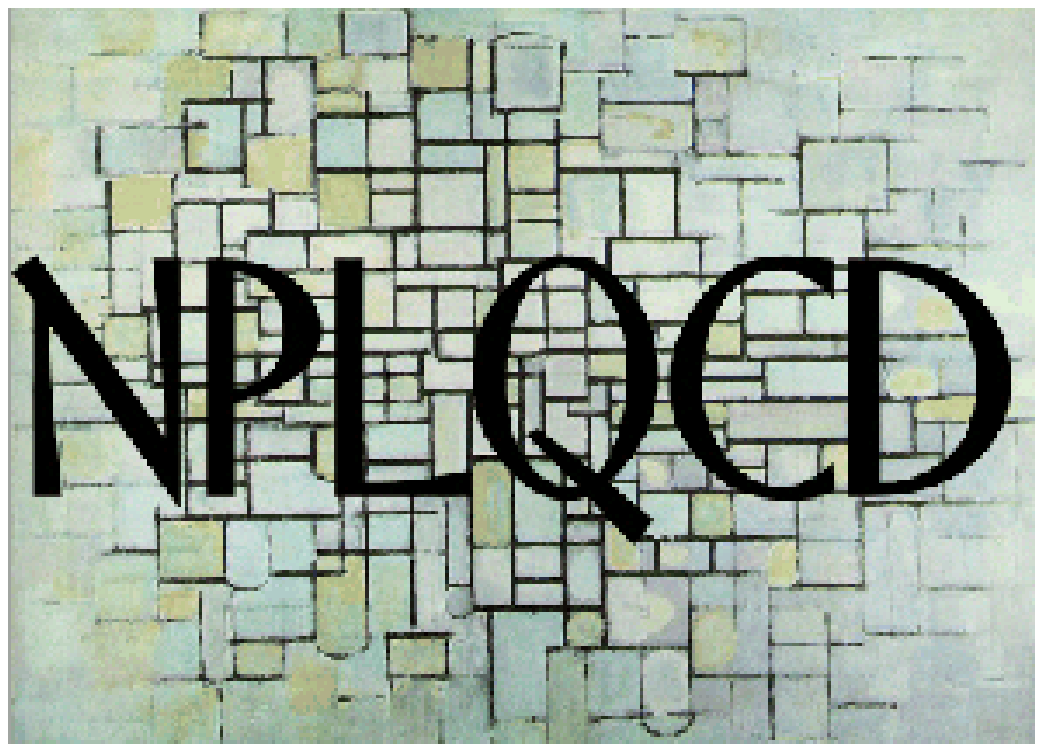}}}
\end{figure}

\preprint{\vbox{
\hbox{NT@UW-05-009}
\hbox{CALT-68-2571}
}}
\vphantom{}

\title{Extrapolation Formulas for Neutron EDM Calculations in Lattice QCD}
\author{Donal O'Connell}
\affiliation{
Department of Physics, California Institute of Technology,
Pasadena, CA 91125, USA.}
\author{Martin J. Savage}
\affiliation{
Department of Physics, University of Washington, 
Seattle, WA 98195-1560, USA.}

\vphantom{}
\vskip 0.5cm
\begin{abstract} 
\vskip 0.5cm
\noindent 
Lattice QCD is rapidly progressing toward being 
able to reliably compute the electric dipole moment of the neutron as 
a function of the strong CP-violating parameter $\overline{\theta}$.
Present day calculations are performed at unphysical values of the light quark masses, 
in volumes that are 
not exceptionally large and at lattice spacings that are not exceptionally small.
We use chiral perturbation theory to determine the leading contributions to the neutron 
electric dipole moment
at finite volume, and in partially-quenched calculations.

\vskip 8.0cm
\leftline{August  2005}
\end{abstract}

\maketitle

\vfill\eject

\section{Introduction}

\noindent
CP-violation is still a mystery.
Current measurements of CP-violating processes in the kaon and B-meson sectors would suggest that
the single phase in the CKM matrix provides a complete description.
However, the baryon asymmetry of the universe cannot be described by this phase alone, and 
there are additional sources of CP-violation that await discovery.
The recent revelation that neutrinos have non-zero masses
has presented us with the possibility of CP-violation in the 
lepton sector. With both Dirac and Majorana type masses possible, 
CP-violation in the neutrino sector is likely to be far more intricate than in the 
quark sector.
The significant number of experiments operating in, and planned to explore the 
neutrino sector will greatly improve our knowledge in this area in the not so distant future.
It has been a puzzle for many years that there is the possibility of strong CP-violation 
arising from the $\theta$ term in the strong interaction sector,
but that there is no evidence at this point in time for 
such an interaction.  
The naive estimate for the size of observables, such as the electric dipole moment 
(edm) of the neutron,  induced by such an interaction is orders of magnitude  
larger than current experimental upper bounds, 
thereby placing a stringent upper bound on the coefficient of the interaction, 
$\overline{\theta}$.
An anthropic argument that compels $\overline{\theta}$ 
to be small does not yet exist and so it is likely
that there is an underlying mechanism, such as the Peccei-Quinn mechanism and 
associated axion, that eliminates this operator.
With the increasingly precise experimental efforts to observe the neutron edm~\cite{brad,Harris:1999jx}, 
it is important to have  a rigorous calculation directly from QCD.

Lattice calculations of the neutron 
edm~\cite{Aoki:1989rx,Aoki:1990ix,Guadagnoli:2002nm,Berruto:2004cr,Shintani:2005xg,TBlumlatt05} 
in terms of the strong CP-violating parameter
are continually evolving toward a reliable 
estimate that can be directly compared with experimental limits and possible future 
observations.
The latest generation of lattice 
calculations respect chiral symmetry and lattice spacing effects have been relegated 
to ${\cal O}(a^2)$. However, the calculations are performed in modest finite volumes and 
at quark masses that are larger than those of nature.
In this work we explore the impact of finite volume on such calculations and also examine 
the quark mass dependence of partially-quenched calculations.

The QCD Lagrange density in the presence of the CP-violating  $\theta$-term is
\begin{eqnarray}
{\cal L} & = & 
\overline{q}\  i\Dslash \ q \ -\ 
\overline{q}_L \ m_q\ q_R
\ -\ 
\overline{q}_R \ m_q^\dagger \ q_L
\ -\ 
{1\over 4}G^{(A)\mu\nu}G^{(A)}_{\mu\nu}
\ +\ 
\theta {g^2\over 32\pi^2} G^{(A)}_{\mu\nu} \tilde G^{(A)\mu\nu}
\ \ \ ,
\label{eq:qcd}
\end{eqnarray}
where $\tilde G^{(A)\mu\nu} = {1\over 2} \varepsilon^{\mu\nu\alpha\beta} G^{(A)}_{\alpha\beta}$, $\varepsilon^{0123} = +1$,
and where $q=(u,d)^T$ for two-flavor QCD.
Chiral redefinitions of the quark fields modify the coefficient of the $G\tilde G$ operator 
through the strong anomaly, and as a consequence
it is the quantity $\overline{\theta}=\theta- {\rm arg}({\rm det}(m_q))$ that has physical meaning.
For our purposes it is convenient to start with a Lagrange density where $m_q$ 
is real and diagonal, and 
$\theta$ in eq.~(\ref{eq:qcd}) is equal to $\overline{\theta}$.
One can then remove the $G\tilde G$ operator by a chiral transformation, 
$q_{j R}\rightarrow e^{i\phi_j/2} q_{j R}$, and $q_{j L}\rightarrow e^{-i\phi_j/2} q_{j L}$
subject to the constraint that $\overline{\theta}=-\sum \phi_j$.
Under this transformation the elements of the quark mass matrix become $m_j\rightarrow m_j e^{i \phi_j} $. 

The low-energy effective field theory (EFT) describing the behavior of the pseudo-Goldstone 
bosons 
associated with the breaking of chiral symmetry is, at leading order,
\begin{eqnarray}
{\cal L} & = & {f^2\over 8} {\rm Tr}\left[\ D^\mu\Sigma\ D_\mu\Sigma^\dagger\ 
\right]
\ +\ \lambda\ {f^2\over 4}\ {\rm Tr}\left[\ m_q \Sigma^\dagger\ +\ m_q^\dagger \Sigma\ \right]
\ \ \ ,
\label{eq:pionsLO}
\end{eqnarray}
where $f\sim 132~{\rm MeV}$ is the pion decay constant,
the covariant derivative describing the coupling of the pions to the
electromagnetic field $\mathcal{A}_\mu$ is $D_\mu \Sigma = \partial_\mu \Sigma + i e [Q, \Sigma] \mathcal{A}_\mu$ with $e > 0$, 
and 
$\Sigma\rightarrow L\Sigma R^\dagger$ under chiral transformations,
\begin{eqnarray}
\Sigma & = & e^{2 i M\over f}
\ \ ,\ \ 
M\ =\ 
\left(
\begin{array}{cc}
\pi^0/\sqrt{2}  & \pi^+  \\
 \pi^-   &  -\pi^0/\sqrt{2}  
\end{array}
\right)
\ \ \ .
\label{eq:sigma}
\end{eqnarray}
We are restricting ourselves to the two-flavor case, but the arguments are general.
In order for the pion field in eq.~(\ref{eq:sigma}) to be a fluctuation about the 
true strong interaction ground state, the phases $\phi_j$ are constrained so 
that in the expansion of eq.~(\ref{eq:pionsLO}), terms linear in the pion field are absent.
The two constraints on the phases lead to the well known relations
\begin{eqnarray}
\phi_u & = & -{\overline{\theta}\  m_d\over m_u+m_d}
\ \ ,\ \ 
\phi_d \ = \ -{\overline{\theta}\  m_u\over m_u+m_d}
\ \ \ ,
\label{eq:phases}
\end{eqnarray}
where we have used the fact that $\overline{\theta}\ll 1$.

\section{Strong CP-Violation in Chiral Perturbation Theory}

\noindent
At leading order in the heavy baryon expansion, the nucleon dynamics are described 
by a Lagrange density 
of the form
\begin{eqnarray}
{\cal L} & = & \overline{N}\ i v\cdot D N\ +\ 2 g_A \overline{N}\ S^\mu A_\mu\  N
\ \ \ ,
\label{eq:LON}
\end{eqnarray}
where $v_\mu$ is the nucleon four-velocity and $S_\mu$ is the covariant spin operator.
The chiral covariant derivative is given in terms of the meson vector field
$D_\mu N = \partial_\mu N + V_\mu N$, where
$V_\mu = {1\over 2}\left( \xi^\dagger\partial_\mu\xi + \xi\partial_\mu\xi^\dagger \right)$.
The field $\xi$ is related to the $\Sigma$-field in eq.(\ref{eq:sigma}) by $\Sigma=\xi^2$,
and $\xi\rightarrow L\xi U^\dagger = U\xi R^\dagger$ under chiral transformations.
The leading order interaction 
between nucleons and the pions is characterized by the axial coupling constant 
$g_A\sim 1.26$ in eq.~(\ref{eq:LON}), where 
$A_\mu = {i\over 2}\left( \xi\partial_\mu\xi^\dagger - \xi^\dagger\partial_\mu\xi \right)$.
The light quark masses contribute to the dynamics of nucleons through the Lagrange density
\begin{eqnarray}
{\cal L}_m & = & -\ 2 \alpha\ \overline{N}\ m_{q\xi+}\ N 
\ - 2 \ \sigma\ \overline{N}\ N\ {\rm Tr}\left(\ m_{q\xi+}\ \right)
\ \ \ ,
\label{eq:Nmass}
\end{eqnarray}
where 
$m_{q\xi+}={1\over 2}\left( \xi^\dagger m_q\xi^\dagger + \xi m_q^\dagger \xi\ \right)$,
$m_q\rightarrow L m_q R^\dagger$, and $m_{q\xi+}\rightarrow U m_{q\xi+} U^\dagger$.
The quantities $\alpha$ and $\sigma$ are constants that must be determined experimentally.
Upon removing the $G\tilde G$ term by a chiral transformation, the mass matrix becomes
$m_q={\rm diag}\left(m_u e^{i\phi_u} , m_d e^{i\phi_d}\right)$, where 
$\phi_{u,d}$ are given in eq.~(\ref{eq:phases}).
Neglecting electromagnetic contributions to the nucleon mass splitting, and using  
light quark masses of 
$m_u=5~{\rm MeV}$ and $m_d=10~{\rm MeV}$, we find 
that~\footnote{The standard analysis usually invokes the approximate flavor 
$SU(3)$ symmetry, e.g. Ref.~\protect\cite{Pich:1991fq}.
The light quark contribution to the baryon masses arises from
\begin{eqnarray}
{\cal L}_m & = & -b_0\ {\rm Tr}\left[ \ m_{q\xi+}\ \right] 
{\rm Tr}\left[\ \overline{B}\  B\ \right] 
\ -\ 
b_1\ {\rm Tr}\left[\ \overline{B}\  m_{q\xi+}\ B \right]
\ -\ 
b_2\ {\rm Tr}\left[\ \overline{B}\ B\ m_{q\xi+}\ \right]
\ \ \ ,
\end{eqnarray}
from which one finds that, neglecting electromagnetic corrections,
\begin{eqnarray}
b_1 & = & {M_{\Xi^0}-M_{\Sigma^+}\over m_s-m_u}\ \sim\ 1.1
\ \ ,\ \ 
b_2 \ = \ {M_p- M_{\Sigma^+}\over m_s-m_d}\ \sim\ -2.3
\ \ \ ,
\end{eqnarray}
where we have used $m_s\sim 120~{\rm MeV}$.
}
\begin{eqnarray}
2 \alpha = {M_p-M_n\over m_u-m_d}\ \sim\ 0.26
\ \ \ .
\label{eq:alval}
\end{eqnarray}
The sigma term is defined to be 
\begin{eqnarray}
\sigma_N=\sum_{u,d} m_q {\partial M_N\over \partial m_q}
\ \ \ ,
\end{eqnarray}
where $M_N$ is the nucleon mass in the isospin limit $m_u, m_d \rightarrow
\overline m$, and is related to the quantities $\alpha$ and $\sigma$ by
\begin{equation}
\sigma_N = 2 (\alpha + 2 \sigma) \overline m .
\end{equation}
The value of $\sigma_N$ is somewhat uncertain, with values ranging
between $45\pm 8~{\rm MeV}$~\cite{Gasser:1990ce} and $64\pm 7~{\rm
MeV}$~\cite{Pavan:2001wz}. Partially-quenched lattice computations are
currently underway to evaluate both $\alpha$ and $\sigma_N$.

Expanding out the interaction in eq.~(\ref{eq:Nmass}) to linear order in the pion 
field gives rise to the CP-violating, momentum independent $NN\pi$ vertex
\begin{eqnarray}
{\cal L} & = & -{4\ \alpha\ \overline{\theta}\over f} \ {m_u\  m_d\over m_u+m_d} 
\ \overline{N}\ 
\left(
\begin{array}{cc}
\pi^0/\sqrt{2}  & \pi^+  \\
 \pi^-   &  -\pi^0/\sqrt{2}  
\end{array}
\right)
\ N
\ +\ ...
\ \ \ .
\label{eq:cpviol}
\end{eqnarray}
It is well known that a
single insertion of this interaction into a one-loop diagram  gives rise to 
an electric dipole moment (edm) of the 
nucleon~\cite{Crewther:1979pi}~\footnote{This set of diagrams also dominates
  the nucleon edm form-factor, as recently computed in Ref.~\cite{HvK}.}.
\begin{figure}[!ht]
\centerline{{\epsfxsize=4.0in \epsfbox{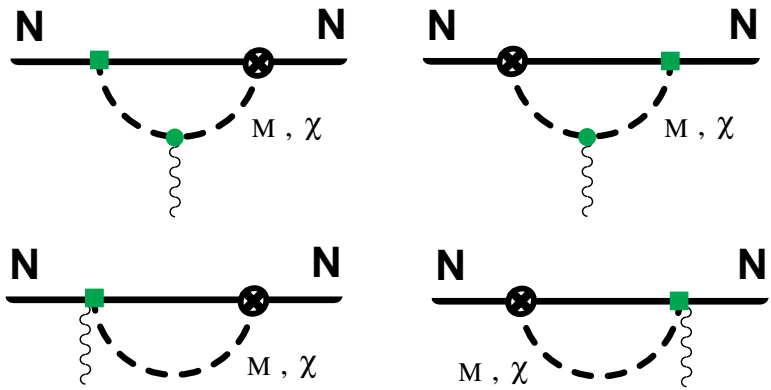}}} 
\vskip 0.15in
\noindent
\caption{\it 
The one-loop diagrams that contribute to the neutron edm in chiral
perturbation theory.
In QCD only $\pi$'s participate in the loop diagram, while in partially-quenched QCD
there are contributions from the bosonic mesons, $M$, and the fermionic mesons, $\chi$.
The crossed circle denotes an insertion of the CP-violating vertex in 
eq.~(\protect\ref{eq:cpviol}), the square denotes an insertion of the strong 
$\pi$NN or $\pi \gamma$NN interaction from eq.~(\protect\ref{eq:LON}) with
derivatives promoted to covariant derivatives,
and the small circle denotes an electromagnetic interaction 
with the meson from eq.~(\protect\ref{eq:pionsLO}) .
}
\label{fig:loops}
\vskip .2in
\end{figure}

The electric dipole moment of the neutron, $d_n$,  is defined by the Lagrange density
describing the interaction between a neutron and an external electric field,
\begin{eqnarray}
{\cal L} & = & d_n\ \overline{n}\  {\boldsymbol \sigma}\cdot {\bf E}\  n
\ \ \ ,
\end{eqnarray}
where ${\boldsymbol \sigma}$ are the Pauli spin matrices,
and ${\bf E}$ is an external electric field.
A calculation of the one-loop diagrams shown in Fig.~\ref{fig:loops} leads to 
\begin{eqnarray}
d_n & = & {g_A\ \alpha\ e\ \overline{\theta}\over 2 \pi^2 f^2} {m_u \ m_d\over m_u+m_d}
\log\left({m_\pi^2\over\mu^2}\right)\ +\ 
\overline{\theta}\ {m_u \ m_d\over m_u+m_d}\ {e\over\Lambda_\chi^2}\ c(\mu) 
\ \ \ ,
\label{eq:nedm}
\end{eqnarray}
We have only kept the logarithmic contribution from the loop diagram, 
which depends upon the 
renormalization scale $\mu$.  This scale dependence is exactly compensated 
by the contributions from local counterterms, which we have combined into $c(\mu)$.
There are ten local counterterms that 
contribute to the nucleon edm, as presented in Ref.~\cite{Pich:1991fq},
and setting $\mu\sim\Lambda_\chi$ we anticipate that $c(\Lambda_\chi)\sim 1$,
where $\Lambda_\chi\sim 1~{\rm GeV}$ is the scale of chiral symmetry breaking.

Numerically, using the value of $\alpha$ in eq.~(\ref{eq:alval}), 
we find the one-loop contribution to be 
\begin{eqnarray}
d_n & \sim & -1.2 \times 10^{-16}\ \overline{\theta}\ \ \ \ {\rm e\ cm}
\ \ \ ,
\end{eqnarray}
which is consistent with previous estimates~\cite{Crewther:1979pi,Pich:1991fq} of the one-loop diagram~\footnote{In Refs.~\cite{Crewther:1979pi,Pich:1991fq} the
electronic charge $e$ is negative.}.
The current experimental upper limit is $|d_n| < 6.3\times 10^{-26} \ \ {\rm e\ cm}$,
from which one concludes that $| \overline{\theta} | \lsim 5\times 10^{-10}$.

\section{Neutron EDM at Finite Volume}

\noindent
Lattice calculations of the neutron edm are performed on finite lattices, 
and therefore one must consider finite volume corrections in translating the lattice results 
to physical predictions.  For large enough lattices, of course, 
these finite volume effects will be exponentially small~\cite{Luscher:1985dn,Luscher,Beane:2004tw}.
There has been a fair amount of work on finite volume corrections to quantities 
calculated in lattice QCD~\cite{Gasser:1987zq,betterways,Luscher,Colangelo:2002hy,Colangelo:2003hf,S92,golter1,Pqqcd2,golter3,davidlin,Becirevic:2003wk,ArLi,Colangelo:2004xr,AliKhan:2002hz,AliKhan:2003kb,AliKhan:2003rw,Khan:2003cu,Kronfeld:2002pi,Beane:2004tw,Beane:2004rf,Young:2004tb,Thomas:2005qm},
but it is only recently that the properties of baryons, and in particular the nucleon, have been
considered~\cite{Beane:2004tw,Beane:2004rf,Young:2004tb,Thomas:2005qm}.

In the calculations that follow, we will assume that the time direction of the lattice 
is infinite.  Clearly, this can only be an approximation, but in most simulations, 
the time direction is considerably larger than the spatial directions, usually
by more than a factor of two.
By assuming that it is infinitely long, we are able to analytically 
perform the integral over energy in 
the loop diagrams that contribute to the observable of interest, 
leaving sums over the allowed three-momentum modes on the lattice.
Details of this procedure can be found in Refs.~\cite{Beane:2004tw,Beane:2004rf}, 
and we will not elaborate further here.
By computing the one-loop diagrams in Fig.~\ref{fig:loops} 
in a finite spatial volume for which the 
spatial dimension, $L$, is much greater than the pion Compton wavelength, $m_\pi L \gg 1$,
and for which the power counting rules are those of the p-regime at infinite volume, 
we find that
\begin{eqnarray}
d_n^{(L)} & = & d_n^{(\infty)}\ -\ 
{g_A\ \alpha\ e\ \overline{\theta}\over \pi^2 f^2} {m_u \ m_d\over m_u+m_d}
\ \sum_{{\bf n}\ne {\bf 0}}\ K_0\left(m_\pi L |{\bf n}|\right)
\ \ \ ,
\end{eqnarray}
where $d_n^{(L)}$ is the neutron edm at finite volume, and $d_n^{(\infty)}$ is
its value at infinite volume.  $K_0(x)$ is a modified Bessel function.
\begin{figure}[!ht]
\centerline{{\epsfxsize=4.0in \epsfbox{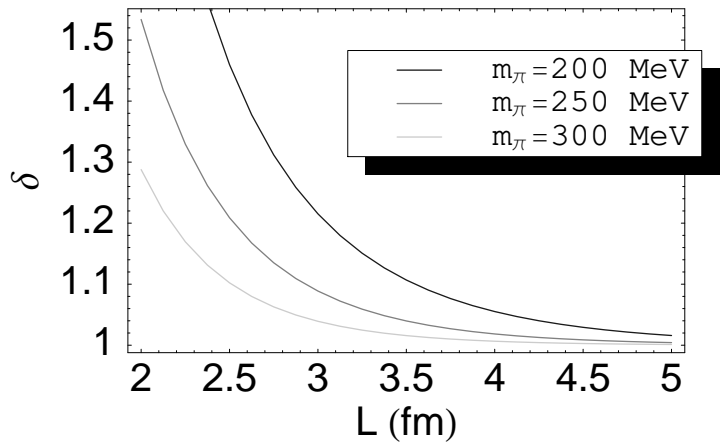}}} 
\vskip 0.15in
\noindent
\caption{\it 
The ratio, $\delta = d_n^{(L)}/d_n^{(\infty)}$,  
of the neutron edm at finite volume to its value at
infinite volume as a function of spatial lattice size $L$, for $c(\mu)=0$.
}
\label{fig:p-regime}
\vskip .2in
\end{figure}
In Fig.~\ref{fig:p-regime} we show the ratio $d_n^{(L)}/d_n^{(\infty)}$ for $c(\mu)=0$,
as an example.  The finite volume corrections are found to be quite large, primarily 
due to the 
fact that the leading order contribution to the edm is at the one-loop level, and not 
from a lower dimension operator.
In the case of the
nucleon properties previously considered~\cite{Beane:2004tw,Beane:2004rf}, 
such as the nucleon mass, magnetic moment and 
matrix elements of the axial current, the loop contributions are subleading, and hence the 
finite volume corrections are subleading in the effective field theory.

As one moves into smaller volumes, where $m_\pi L\lsim 1$, the p-regime power counting 
is no longer applicable, and we move into the $\epsilon^\prime$-regime~\cite{Detmold:2004ap}.
In this regime, the spatial zero-modes are enhanced relative to the non-zero-modes, and a 
power counting in terms of the small parameter  $\epsilon^\prime=m_\pi L$ is appropriate.
For the neutron edm calculation, the same one-loop diagram shown in 
Fig.~\ref{fig:loops} makes the leading contribution in $\epsilon^\prime=m_\pi L$,
and the neutron edm is found to be 
\begin{eqnarray}
d_n^{(L)} & = & 
{-2\ g_A\ \alpha\ e\ \overline{\theta}\over f^2\ m_\pi^3\ L^3} {m_u \ m_d\over m_u+m_d}
\ +\ \cdots
\ \ \ ,
\end{eqnarray}
where the ellipses denote terms higher order in the $\epsilon^\prime$-expansion.
The classic exponential behavior of the p-regime becomes power law, $1/L^3$, behavior
as the spatial volume decreases.

\section{Neutron EDM in Partially-Quenched QCD}

\noindent
While partially-quenched QCD (PQQCD)~\cite{BG94,Sh97,GL98,SS00,SS01,LS96,Sharpe:2001fh}
is not a theory that describes nature, it is a theory
that can be used to describe unphysical lattice calculations, and allows the direct 
extraction of QCD observables via an extrapolation in quark-masses.
In calculating quantities in lattice QCD, the quark masses used in the generation of
gauge field configurations does not have to be the same as the quark masses of the propagators
computed on those configurations.
The reason why this is a useful concept is that the computer time required to generate a 
dynamical configuration grows rapidly as the quark mass is reduced, while the time to compute 
a propagator grows more slowly. 
Currently, lattice calculations cannot be performed at the physical quark masses, 
but we wish to be as ``close as possible'' to the physical values in order to minimize the 
impact of quark mass extrapolations.

The Lagrange density describing the quark sector of PQQCD is
\begin{eqnarray}
{\cal L} & = & 
\sum_{k,n=u,d,\tilde u, \tilde d,j,l} 
\overline{Q}^k\ 
\left[\ i\Dslash -m_{Q}\ \right]_k^n\ Q_n
\ -\ 
{1\over 4}G^{(A)\mu\nu}G^{(A)}_{\mu\nu}
\ +\ 
\theta {g^2\over 32\pi^2} G^{(A)}_{\mu\nu} \tilde G^{(A)\mu\nu}
\ \ \ ,
\label{eq:PQQCD}
\end{eqnarray}
where the left- and right-handed 
valence, sea, and ghost quarks are combined into
column vectors
\begin{eqnarray}
Q_L & = & \left(u,d,j,l,\tilde u,\tilde d\right)^T_L
\ \ ,\ \ 
Q_R \ =\  \left(u,d,j,l,\tilde u,\tilde d\right)^T_R
\ \ \ .
\label{eq:quarkvec}
\end{eqnarray}
The objects $\eta_k$ correspond to the parity of the component of $Q_k$,
with $\eta_k=+1$ for $k=1,2,3,4$ and $\eta_k=0$ for $k=5,6$.
The $Q_{L,R}$ in eq.~(\ref{eq:quarkvec})
transform in the fundamental representation of $SU(4|2)_{L,R}$
respectively.
The ground floor of $Q_L$ transforms as a $({\bf 4},{\bf 1})$ of
$SU(4)_{qL}\otimes SU(2)_{\tilde q L}$ while the first floor transforms
as $({\bf 1},{\bf 2})$, and the 
right handed field $Q_R$ transforms analogously.
In the absence of quark masses, $m_Q=0$,
the Lagrange density in eq.~(\ref{eq:PQQCD})
has a
graded symmetry $U(4|2)_L\otimes U(4|2)_R$, where the left- and 
right-handed quark fields transform as
$Q_L\rightarrow U_L Q_L$ and $Q_R\rightarrow U_R Q_R$ respectively.
The strong anomaly reduces the symmetry of the theory, which can be
taken to be 
$SU(4|2)_L\otimes SU(4|2)_R\otimes U(1)_V$~\cite{SS01}.
It is assumed that this symmetry is spontaneously broken 
$SU(4|2)_L\otimes SU(4|2)_R\otimes U(1)_V\rightarrow 
SU(4|2)_V\otimes U(1)_V$ so that an identification with QCD can be made.
The mass matrix, $m_Q$, has entries
$m_Q = {\rm diag}(m_u,m_d,m_j,m_l,m_u,m_d)$,
(i.e.  the valence quarks and ghosts are degenerate)
so that the contribution 
to the determinant in the path integral from 
the valence quarks and ghosts exactly cancel, leaving the 
contribution from the sea quarks alone.
This makes clear why the partially-quenched theory describes lattice calculations with
sea quarks and valence quarks of differing mass.
The details concerning the construction of partially-quenched chiral perturbation theory 
(PQ$\chi$PT) are well known and can be found in several works, 
e.g. Ref.~\cite{Chen:2001yi,Beane:2002vq}.
The quantity that has ``physical'' impact for lattice calculations 
of strong CP-violating quantities is~\footnote{
\begin{eqnarray}
{\rm sdet}\left(
\begin{array}{cc}
A & B\\
C & D
\end{array}
\right)
& = & {{\rm det}\left(A- B D^{-1} C\right)\over {\rm det}\left(D\right)}
\ \ \ .
\end{eqnarray}
}
$\overline{\theta}=\theta- {\rm arg}\left( {\rm sdet}\left(m_Q\right)\right)$.

The strong interaction dynamics of the 
pseudo-Goldstone bosons are described  at leading order (LO)
in PQ$\chi$PT by a Lagrange density of the form,
\begin{eqnarray}
{\cal L } & = & 
{f^2\over 8} 
{\rm str}\left[\ \partial^\mu\Sigma^\dagger\partial_\mu\Sigma\ \right]
 \ +\ 
\lambda\ {f^2\over 4} 
{\rm str}\left[\ m_Q\Sigma^\dagger + m_Q^\dagger\Sigma\ \right]
\ +\ 
\alpha_\Phi\partial^\mu\Phi_0\partial_\mu\Phi_0\ -\ m_0^2\Phi_0^2
 \ ,
\label{eq:lagpi}
\end{eqnarray}
where $\alpha_\Phi$ and $m_0$ are quantities that do not vanish in the 
chiral limit.
In order to simply project out the singlet of the graded group one takes 
the limit $m_0\rightarrow\infty$~\cite{Sharpe:2001fh}.
The meson field is incorporated in $\Sigma$ via
\begin{eqnarray}
\Sigma & = & \exp\left({2\ i\ \Phi\over f}\right)
\ =\ \xi^2
\ \ \ ,\ \ \ 
\Phi \ =\  \left(\matrix{ M &\chi^\dagger \cr \chi &\tilde{M} }\right)
\ \ \ ,
\label{eq:phidef}
\end{eqnarray}
where $M$ and $\tilde M$ are matrices containing bosonic mesons while
$\chi$ and $\chi^\dagger$ are matrices containing fermionic mesons,
with
\begin{eqnarray}
M & = & 
\left(\matrix{
\eta_u & \pi^+ & J^0 & L^+ \cr
\pi^- & \eta_d & J^- & L^0\cr
\overline{J}^0 & J^+ & \eta_j & Y_{jl}^+\cr
L^- & \overline{L}^0 & Y_{jl}^- & \eta_l  }
\right)
,
\tilde M  =  \left(\matrix{\tilde\eta_u & \tilde\pi^+ \cr
\tilde\pi^- & \tilde\eta_d
}\right)
,
\chi =
\left(\matrix{\chi_{\eta_u} & \chi_{\pi^+} &  
\chi_{J^0} & \chi_{L^+}\cr
\chi_{\pi^-} & \chi_{\eta_d} & 
\chi_{J^-} & \chi_{L^0} }
\right)
,
\label{eq:mesdef}
\end{eqnarray}
where the upper $2\times 2$ block of $M$ is the usual triplet plus singlet 
of pseudo-scalar mesons while the remaining entries correspond to mesons
formed from the sea quarks.
The convention we use corresponds to $f~\sim~132~{\rm MeV}$,
and the charge assignments have been made using an electromagnetic charge
matrix of ${\cal Q}^{(PQ)} = {1\over 3} {\rm diag}\left(2,-1,2,-1,2,-1\right)$.
For the calculations we will be performing, the flavor singlet pseudo-Goldstone 
boson does not contribute, and so we do not discuss it and its associated hairpin interactions.

The free Lagrange density describing the interactions of the nucleon and its superpartners
which are embedded in the  ${\bf 70}$ dimensional irreducible representation of $SU(4|2)$
${\cal B}_{ijk}$ is, at LO in the heavy baryon 
expansion~\cite{JMheavy,JMaxial,Jmass,chiralN},
\begin{eqnarray}
{\cal L} & = & 
i\left(\overline{\cal B} v\cdot {\cal D} {\cal B}\right)
-2\alpha_M^{\rm (PQ)} \left(\overline{\cal B}{\cal B}{\cal M}_+\right)
-2\beta_M^{\rm (PQ)} \left(\overline{\cal B}{\cal M}_+{\cal B}\right)
-2\sigma_M^{\rm (PQ)} \left(\overline{\cal B}{\cal B}\right)\ 
{\rm str}\left({\cal M}_+\right)
\ ,
\label{eq:free}
\end{eqnarray}
where
${\cal M}_+={1\over 2}\left(\xi^\dagger m_Q\xi^\dagger + \xi m_Q^\dagger\xi\right)$.
The brackets $\left(\ ...\ \right)$ denote contraction of Lorentz and flavor
indices as defined in Ref.~~\cite{LS96}.

The Lagrange density describing the interactions of the 
${\bf 70}$ with the pseudo-Goldstone bosons at LO in the chiral expansion
is~\cite{LS96},
\begin{eqnarray}
{\cal L} & = & 
2\rho\ \left(\overline{\cal B} S^\mu {\cal B} A_\mu\right)
\ +\ 
2\beta\ \left(\overline{\cal B} S^\mu A_\mu {\cal B} \right)
\ ,
\label{eq:ints}
\end{eqnarray}
where $S^\mu$ is the covariant spin vector~\cite{JMheavy,JMaxial,Jmass}.
Restricting ourselves to the valence sector, we can compare eq.~(\ref{eq:ints}) with the 
LO interaction Lagrange density of QCD,
\begin{eqnarray}
{\cal L} & = & 
2 g_A\  \overline{N} S^\mu  A_\mu N
\ +\ 
g_1\overline{N} S^\mu N\ {\rm tr}
\left[\ A_\mu\ \right]
\ ,
\label{eq:intsQCD}
\end{eqnarray}
and find that at tree level
\begin{eqnarray}
\rho & = & {4\over 3} g_A\ +\ {1\over 3} g_1
\ \ \ ,\ \ \ 
\beta \ =\ {2\over 3} g_1 - {1\over 3} g_A
\ \ \ .
\label{eq:axrels}
\end{eqnarray}

The contribution to the strong anomaly from the valence quarks is exactly canceled by the 
contribution from the ghosts.
Therefore, chiral transformations of the sea quarks alone remove the
$\theta$-term from the Lagrange density in eq.~(\ref{eq:PQQCD}).
Upon a chiral transformation of the valence quark, sea quark and ghost fields, 
the quark super-mass matrix becomes
$m_Q = {\rm diag}( m_u e^{i\phi_u} , m_d  e^{i\phi_d} , m_j e^{i\phi_j} , 
m_l e^{i\phi_l} , m_u e^{i\phi_u}  , m_d e^{i\phi_d} )$ 
subject to the constraint that $\overline{\theta}=-\sum \ (-)^{\eta_k+1}\ \phi_k$.
The vacuum stability condition for small $\overline{\theta}$ further provides the constraint 
$ m_u \phi_u = m_d \phi_d = m_j \phi_j = m_l\phi_l$.
Therefore, we have
\begin{eqnarray}
\phi_j & = & -{\overline{\theta} \ m_l\over m_j+m_l}
\ , \ 
\phi_l  =  -{\overline{\theta}\ m_j\over m_j+m_l}
\  , \ 
\phi_u  = -{\overline{\theta}\ m_j\ m_l\over m_j+m_l}\ {1\over m_u}
\  , \ 
\phi_d  = -{\overline{\theta}\ m_j\ m_l\over m_j+m_l}\ {1\over m_d}
\ .
\label{eq:pqangs}
\end{eqnarray}
By using this phase-rotated mass matrix in the Lagrange density of eq.~(\ref{eq:free}),
one induces a CP-violating interaction between the pseudo-Goldstone bosons and 
the baryons of the partially-quenched theory, in precisely the same way as in QCD.
Further, this vertex generates the leading contribution to the neutron edm through the one-loop diagrams analogous to those in 
Fig.~\ref{fig:loops}.
One further slight complication that can be considered is that the electric charge matrix
in the partially-quenched theory is not specified by nature; all that is required is that 
one reproduces QCD in the limit that the sea and valence quarks become 
degenerate~\footnote{Even this constraint is excessive.
It is sufficient 
to determine matrix elements of operators transforming in the  singlet and 
adjoint representations of the graded group.
}~\cite{Golterman:2001qj,Golterman:2001yr,Chen:2001yi,Beane:2002vq}.
In our computations, we use an electric charge matrix of the form
${\cal Q^{(PQ)}}={\rm diag}\left({2\over 3},-{1\over 3},q_j,q_l,q_j,q_l\right)$.

Working in the isospin limit where $m_j=m_l=m_{\rm sea}$, and defining 
$q_{jl}=q_j+q_l$, we find that the leading order contribution to the neutron edm is
\begin{eqnarray}
d_n^{(PQ)} & = & 
{e\ \overline{\theta}\ m_{\rm sea}\over 4\pi^2 f^2}
\left[\ 
F_\pi\ \log\left({m_\pi^2\over \mu^2}\right)
\ +\ 
F_J\ \log\left({m_J^2\over \mu^2}\right)
\ \right]
\nonumber\\
& & \ +\ 
\overline{\theta}\ {e\over\Lambda_\chi^2}\ 
\left[\ {m_{\rm sea}\over 2}\  c(\mu) 
\ +\ d \left(m_{\rm sea}-m_{\rm val}\right)
\ +\ f q_{jl}\ \left(m_{\rm sea}-m_{\rm val}\right)\right]
\ \ \ ,
\label{eq:pqnedm}
\end{eqnarray}
where $m_J$ is the mass of the Goldstone boson composed of a sea quark and a
valence quark, and 
\begin{eqnarray}
 F_\pi & = & g_A \left({2 \alpha_M^{\rm (PQ)} - \beta_M^{\rm (PQ)}\over 3}\right)
- 
g_A \alpha_M^{\rm (PQ)}\left( {1\over 3} + {q_{jl}\over 2}\right)
+ 
g_1 \left( {\beta_M^{\rm (PQ)}\over 3}- 
 \left({\alpha_M^{\rm (PQ)} + 2 \beta_M^{\rm (PQ)}\over 4}\right)q_{jl}
\right)
\nonumber\\
F_J & = & g_A \alpha_M^{\rm (PQ)}\left( {1\over 3} + {q_{jl}\over 2}\right)
\ -\ 
g_1 \left( {\beta_M^{\rm (PQ)}\over 3}\ -\ 
 \left({\alpha_M^{\rm (PQ)} + 2 \beta_M^{\rm (PQ)}\over 4}\right)q_{jl}
\right)
\ \ \ .
\label{eq:Fdef}
\end{eqnarray}
As we can make the tree level identification
$\alpha= (2 \alpha_M^{\rm (PQ)}-\beta_M^{\rm (PQ)})/3$, the expression in 
eq.~(\ref{eq:pqnedm}) reduces to the QCD result in eq.~(\ref{eq:nedm}) 
when $m_{\rm sea}\rightarrow m_{\rm valence}$ and $m_J\rightarrow m_\pi$, 
since $F_\pi + F_J = g_A \alpha$.
It is important to notice that the counterterm that contributes in the partially-quenched case,
$c(\mu)$, is the same as in the QCD case, while the other two counterterms, $d$ and $f$, 
make a vanishing contribution in QCD.
The expression in eq.~(\ref{eq:pqnedm})
exhibits one of the well known pathologies of the partially 
quenched theory.
One sees that this expression behaves as 
$\sim m_{\rm sea}\log \left(m_{\rm valence}\right)$.
For a fixed sea quark mass, the one-loop contribution diverges as the valence 
quarks move toward the chiral limit, in contrast to the case of QCD where the 
diagram diminishes as $\sim m_\pi^2\log \left(m_\pi^2\right)$.

The finite volume corrections resulting from a partially-quenched calculation are 
obviously more complicated than in QCD.
In the limit where the volume is large compared to the Compton wavelength of both the 
valence and sea mesons, one can use the power counting of the p-regime to find that
\begin{eqnarray}
& & d_n^{(PQ)(L)} \ = \ 
d_n^{(PQ)(\infty)}
 \ -\ 
{e\ \overline{\theta}\ m_{\rm sea}\over 2\pi^2 f^2}
\left[\ 
F_J\ S_J\ +\ F_\pi\ S_\pi
\right]
\nonumber\\
& & S_\pi \ = \  \sum_{{\bf n}\ne {\bf 0}}\ K_0\left(m_\pi L |{\bf n}|\right)
\ \ \ \  ,\ \ \ \ 
S_J \ =\  \sum_{{\bf n}\ne {\bf 0}}\ K_0\left(m_J L |{\bf n}|\right)
\ \ \ .
\label{eq:pqvol}
\end{eqnarray}

One can imagine performing a calculation of the neutron edm for lattice parameters such that
$m_\pi L \ll 1$ but $m_J L \gsim 1$. Parametrically, we could arrange for 
$m_\pi/\Lambda_\chi \sim \epsilon^{\prime 2}$, $\Lambda_\chi L\sim 1/\epsilon^\prime$ and $m_J/\Lambda_\chi\gsim \epsilon^\prime$.
In such a scenario, the finite volume correction would become
\begin{eqnarray}
& & d_n^{(PQ)(L)}  = -{e\  \overline{\theta}\  m_{\rm sea} \over  f^2\  m_\pi^3\  L^3}\  F_\pi
\ + \cdots\ \ .
\label{eq:pqvolsilly}
\end{eqnarray}
This somewhat bizarre computational set-up allows one to quite dramatically separate
the contributions to the neutron edm, as the leading contribution results from one-loop graphs 
involving pions, and the contribution from mesons involving the sea quarks is suppressed.
However, the sea quarks play a central role via the CP-violating pion-nucleon coupling.
In the more symmetric scenario in which $m_\pi L , m_J L \lsim 1$, the finite volume expression
becomes
\begin{eqnarray}
d_n^{(PQ)(L)} & = &
-{e\ \overline{\theta}\ m_{\rm sea} \over  f^2\  L^3}
\left[\ {F_\pi\over m_\pi^3}\ +\ {F_J\over m_J^3}\ \right]
\ +\cdots \ \ .
\label{eq:pqvolverysilly}
\end{eqnarray}

\section{Conclusions}

\noindent
A non-zero electric dipole moment of the neutron would provide direct evidence for 
time-reversal violation in nature.  
It continues to be the focus of ever more 
precise experimental measurements, and 
the fact that it has not been observed at the present limits of experimental resolution
provides one of the more intriguing puzzles in modern physics.
In this work we have considered how lattice QCD calculations of 
the neutron edm originating from the QCD $\theta$-term,
performed in a finite volume and at unphysical quark masses,
are related to its value in nature.
We have provided explicit formulas 
that allow for the extrapolation from finite volume 
calculations to the infinite volume limit and for the chiral 
extrapolation of partially-quenched calculations.
In order for these formulas to be useful, lattice QCD calculations of both the light quark 
mass dependence of the nucleon mass, $\alpha$, and the neutron edm are required.
With the lattice value of $\alpha$ known with a given precision, the lattice determination
of the neutron edm will then allow for the counterterm $c(\mu)$ to be determined.
Once these constants are computed, the chiral extrapolation  of the neutron edm 
to the physical quark masses, and to infinite volume, is possible.

\bigskip\bigskip

\acknowledgements

We would like to thank Kostas Orginos for interesting discussions on this
subject and for his involvement in the early stages of this work.
MJS would like to thank the CTP at MIT and  the High Energy and Nuclear Theory
groups at Caltech for very kind hospitality during part of this work.
The work of MJS is supported in part by the U.S.~Dept. of Energy under Grant No.~DE-FG03-97ER4014 while the work of DOC is supported in part by the  U.S.~Dept. of Energy under Grant No.~DE-FG03-9ER40701.

\end{document}